\documentclass[epj]{svjour}
\usepackage{graphics}
\newcommand{\beq}{\begin{equation}}
\newcommand{\eeq}{\end{equation}}
\newcommand{\bea}{\begin{eqnarray}}
\newcommand{\eea}{\end{eqnarray}}

\begin{document}
\title{Thermodynamics of the PNJL model}
\author{C. Ratti\inst{1}, S. R\"o\ss ner\inst{2}, M. A. Thaler\inst{2} \and W. 
Weise\inst{2}
}                     
%
%
\institute{ECT*, I-38050 Villazzano (Trento) ITALY and INFN,
Gruppo Collegato di Trento, via Sommarive, I-38050 Povo (Trento) ITALY
\and Physik-Department, Technische Universit\"at M\"unchen, D-85747 Garching,
GERMANY}
\date{Received: date / Revised version: date}
%
\abstract{
QCD thermodynamics is investigated by means of the Polyakov-loop-extended
Nambu Jona-Lasinio (PNJL) model, in which quarks couple simultaneously to the chiral 
condensate and to a background temporal gauge field representing Polyakov
loop dynamics. The behaviour of the Polyakov loop as a function 
of temperature is obtained by minimizing the thermodynamic potential of the 
system. A Taylor series expansion of the pressure is performed. Pressure 
difference and quark number density are then evaluated up to sixth order
in quark chemical potential, and compared to the corresponding lattice data.
The validity of the Taylor expansion is discussed within our model, through a 
comparison between the full results and the truncated ones.
} 
\maketitle
\section{Introduction}
\label{intro}
QCD thermodynamics has been the subject of intense investigations in recent years.
Thanks to lattice simulations, the equation of state of strongly interacting
matter is now at hand as a function of temperature $T$ and in a limited range 
of quark chemical potential $\mu$. Improved multi-parameter 
re-weighting techniques~\cite{Fodor:2002km,Fodor:2001pe}, Taylor 
series expansion methods~\cite{Allton:2002zi,Allton:2003vx,Allton:2005gk} and
analytic continuation from imaginary chemical 
potential~\cite{Laermann1,deForcrand:2003hx,delia1,delia2} provide
lattice data for the pressure, entropy density, quark density and selected
susceptibilities. In response to these lattice simulations, many 
phenomenological models have been proposed
~\cite{Peshier:1995ty,Levai:1997yx,Peshier:1999ww,Szabo:2003kg,Bluhm:2004xn,Bluhm:2004xn2,Pisarski:2000eq,Rebhan:2003wn,Schneider:2001nf,Thaler:2003uz,Drago:2001gd,Ivanov:2004gq,Karsch:2003zq,Rischke:2003mt}, both to give an interpretation
of the available lattice data in terms of effective degrees of freedom, and to
explore those regions of the phase diagram that cannot be reached on the 
lattice yet.

In this context, encouraged by the successful description of $N_c=2$ 
QCD obtained in
the Nambu Jona-Lasinio model~\cite{Ratti:2004ra}, we have investigated recently
full QCD thermodynamics at zero and
finite quark chemical potential in the framework of a Polyakov-loop extended Nambu 
Jona-Lasinio 
(PNJL) model~\cite{Meisinger:1995ih,Meisinger:2001cq,Fukushima:2003fw,Mocsy:2003qw,Megias:2004hj,Ratti1,Ratti2} in which quarks develop 
quasiparticle masses by
propagating in the chiral condensate, while they couple at the same time to a
homogeneous background (temporal) gauge field representing Polyakov loop 
dynamics.

The ``classic'' NJL model incorporates the chiral symmetry of two-flavour QCD 
and its spontaneous breakdown at $T<T_c$. Gluonic degrees of freedom 
are ``integrated out'' and replaced by a local four-point interaction of
quark colour currents. Subsequent Fierz transformations project this 
interaction into various quark-antiquark and diquark channels. The colour
singlet $q{\bar q}$ modes of lowest mass are identified with the lightest
mesons. Pions properly emerge as Goldstone bosons at $T<T_c$. However, the
local $SU(N_c)$ gauge invariance of QCD is now replaced by a global $SU(N_c)$ 
symmetry in the NJL model, so that the confinement property is lost. Consequently,
standard NJL-type models are bound to fail in attempts to describe $N_c=3$
thermodynamics around $T_c$ (and beyond) for non-zero quark chemical potential
$\mu$.

The deconfinement phase transition is well defined in the heavy-quark limit
where the Polyakov loop serves as an order parameter. In the presence of 
dynamical quarks the $Z(3)$ center symmetry of the $SU(3)$ gauge group is explicitly broken.
No rigorous order parameter is established for the deconfinement
transition in this case~\cite{Fukushima:2002bk}, but the Polyakov loop
still serves as an indicator of a rapid crossover towards deconfinement.
 
In the present paper we explore the capability of an updated PNJL model to describe essentials of QCD thermodynamics around and above $T_c$. We use a modified version of our previous PNJL model~\cite{Ratti1,Ratti2}, with an improved effective potential for the Polyakov loop field.   One of our main tasks here is to check the validity of the Taylor expansion approach
at finite chemical potential $\mu$.  The presently available lattice simulations that we want to
compare our results with, provide the equation of state of two-flavor QCD at finite
quark chemical potential by means of a Taylor expansion around $\mu=0$ up to
sixth order in $\mu$. Our PNJL model enables us to produce both the full 
result at finite $\mu$ and the Taylor expansion around $\mu=0$. In this way
we can draw conclusions about the range of applicability of the Taylor 
expansion and its convergence properties.

\section{The model}
\label{sec:1}
The Lagrangian of the Polyakov-Loop-extended Nambu Jona-Lasinio (PNJL) model
is written in the following way:
\bea
\mathcal{L}&=&\bar{\psi}\left(i\gamma_{\mu}D^{\mu}-\hat{m}_0
\right)\psi+\frac
{G}{2}\left[\left(\bar{\psi}\psi\right)^2+\left(\bar{\psi}i\gamma_5
\vec{\tau}\psi
\right)^2\right]
\nonumber\\
&-&\mathcal{U}\left(\Phi,\Phi^*,T\right),
\label{lagr}
\eea
where $\psi=\left(\psi_u,\psi_d\right)^T$ is the quark field,
$\hat{m}_0 = diag(m_0, m_0)$ is the isospin-symmetric bare quark mass matrix
and we define
\beq
D^{\mu}=\partial^{\mu}-iA^{\mu}~~~~~~\mathrm{with}~~~~~~
A^{\mu}=\delta^{\mu}_{0}A^0,~~~~~~A^0=g\mathcal{A}_{a}^{0}\frac{\lambda_a}{2}
\label{def}
\eeq
with $A_{0}=-iA_4$; $\lambda_a$ are the eight Gell-Mann matrices.
 A local, chirally 
symmetric scalar-pseudoscalar four-point interaction of the quark fields is 
introduced with an effective coupling strength $G$.
The potential $\mathcal{U}\left(\Phi,\Phi^*,T\right)$ in the
Lagrangian~(\ref{lagr}) governs the dynamics of the
Polyakov loop $\Phi=Tr_c(L)/3$ and its conjugate $\Phi^*=
Tr_c(L^{\dagger})/3$. The matrix $L$ is written in terms of the
temporal gauge fields:
\beq
L=\left[\mathcal{P}\exp\left(i\int_{0}^{\beta}
A_4d\tau\right)\right]=\exp\left[\frac{iA_4}{T}\right].
\label{l}
\eeq
In a convenient gauge (the so-called Polyakov-gauge) the Polyakov loop
matrix can be given a diagonal representation~\cite{Fukushima:2003fw}.
Note that in the chiral limit ($\hat{m}_0 \rightarrow 0$), this Lagrangian is 
invariant under the chiral flavour group, $SU(2)_L \times SU(2)_R$, just like 
the original QCD Lagrangian. 

The effective potential $\mathcal{U}\left(\Phi,\Phi^*,T\right)$ must obey
the following general features: it must satisfy the $Z(3)$ center symmetry, 
like the pure gauge QCD Lagrangian;
besides, in accordance with lattice predictions for the behaviour of the
Polyakov loop, ${\cal U}$ must have an absolute minimum at $\Phi$=0 at small
temperatures, while above the critical temperature ($T_0\simeq$ 270 MeV 
in pure gauge QCD) the minimum is shifted to a finite value of $\Phi$. In the 
limit $T\rightarrow\infty$ we have $\Phi\rightarrow~1$. 
In our previous works~\cite{Ratti1,Ratti2} we have chosen the simplest possible
form for ${\cal U}$, namely a polynomial in $\Phi,~\Phi^*$. An improved effective potential, 
following Ref.~\cite{Fukushima:2003fw}, replaces the fourth order term 
by $log[J(\Phi)]$, where $J(\Phi)$ is the Jacobi determinant which results from 
integrating out six non-diagonal group elements while keeping the two diagonal ones to represent 
$L$.
We therefore use the following ansatz for $\cal{U}$:
\bea
{\cal U}\left(\Phi,\Phi^*,T\right)&=&-\frac{1}{2}b_2\left(T\right)\Phi^*\Phi
\label{u1}\\
&&\!\!\!\!\!\!\!\!\!\!\!\!\!\!\!\!\!\!+b_4\left(T\right)\log\left[1-6\Phi^*\Phi+4\left({\Phi^*}^3+\Phi^3\right)
-3\left(\Phi^*\Phi\right)^2\right]
\nonumber
\eea
with 
\beq
b_2\left(T\right)=a_0+a_1\left(\frac{T_0}{T}\right)
+a_2\left(\frac{T_0}{T}\right)^2,~~~~~~b_4\left(T\right)=b_4\left(\frac{T_0}{T}
\right)^3.
\label{u2}
\eeq
Through its logarithmic divergence
as $\Phi,~\Phi^*\rightarrow 1$, this ansatz guarantees that the Polyakov loop will be 
automatically constrained to be always smaller than 1, reaching this asymptotic value as $T\rightarrow\infty$.

\begin{figure}
\resizebox{0.45\textwidth}
{!}
{%
  \includegraphics{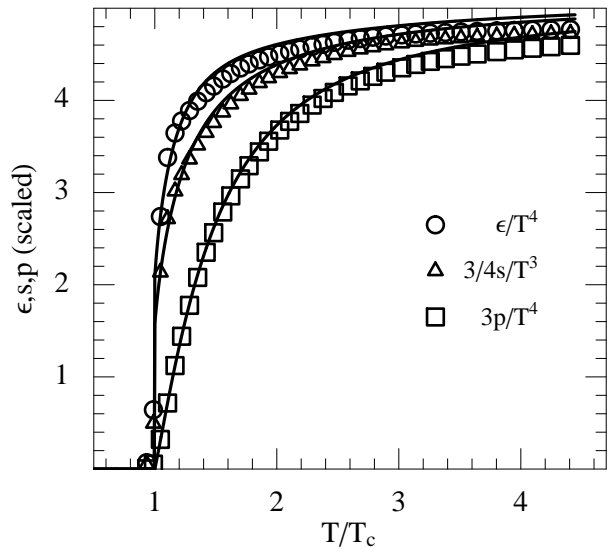}
}
\caption{Scaled pressure, entropy density and energy density as functions
of the temperature in the pure gauge sector, compared to the corresponding 
lattice data taken from Ref.~\cite{Boyd}.}
\label{fig1}       
\end{figure}

\begin{figure}
\resizebox{0.45\textwidth}
{!}
{%
  \includegraphics{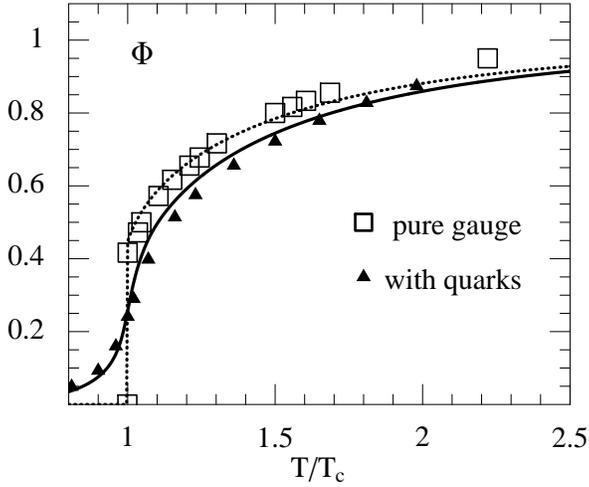}
}
\caption{Dotted line: PNJL fit of Polyakov loop as a function of  
temperature in the pure gauge sector, compared to corresponding lattice 
results (empty symbols) taken from Ref.~\cite{Kaczmarek:2002mc}.
Continuous line: PNJL model prediction of the Polyakov loop behaviour
as a function of the temperature, in the presence of dynamical quarks. The
corresponding lattice data (full symbols) are taken from Ref.~
\cite{Kaczmarek:2005}.}
\label{fig2}       
\end{figure}

Following the procedure of~\cite{Ratti1}, a precision fit of the parameters 
$a_i,~b_i$ is performed to reproduce lattice data for
pure gauge QCD thermodynamics and the behaviour of the Polyakov loop
as a function of temperature.
The results of this combined fit are shown in Figs.~\ref{fig1} and \ref{fig2} 
(dotted line). The corresponding parameters are listed in Table~\ref{tab1}.
The critical temperature $T_0$ for deconfinement in the pure gauge sector is
fixed at 270 MeV, in agreement with lattice results.
\begin{table}
\begin{center}
\begin{tabular}{|c|c|c|c|}
\hline
\hline
&&&\vspace{-.3 cm}\\
$a_0$&$a_1$&$a_2$&$b_4$\\
&&&
\vspace{-.3cm}\\
\hline
\vspace{-.3cm}\\
3.51&-2.47&15.22&-1.75\\
\hline
\hline
\end{tabular}
\caption{Parameter set used in this work for the Polyakov loop 
potential~(\ref{u1},~\ref{u2}).}
\label{tab1}
\end{center}
\end{table}
The NJL part of the model involves some free parameters, which we
take from Ref.~\cite{Ratti1}, where they were fixed to reproduce some
physical quantities in the hadronic sector. These parameters, and the
corresponding physical quantities, are listed in Table~\ref{tab2}.

Before passing to the actual calculations, we summarize basic 
assumptions behind eq.~(\ref{lagr}) and comment on limitations to be kept in 
mind. In fact
the PNJL model~(\ref{lagr}) is quite schematic in several respects. It reduces 
gluon dynamics to a) chiral point couplings between quarks, and b) a simple
static background field representing the Polyakov loop. This
picture cannot be expected to work beyond a limited range of temperatures.
At large $T$, transverse gluons are known to be thermodynamically active 
degrees
of freedom, but they are ignored in the PNJL model. To what extent this model
can reproduce lattice QCD thermodynamics is nonetheless a relevant question. We
can assume that its range of applicability is, roughly, $T\leq (2-3)T_c$, based
on the conclusion drawn in Ref.~\cite{Meisinger:2003id} that transverse gluons
start to contribute significantly for $T>2.5\,T_c$. 

\section{Results}
\subsection{PNJL model at mean-field level}
After performing a bosonization of the PNJL Lagrangian and introducing
the auxiliary fields $\sigma$ and ${\vec\pi}$ the
thermodynamic potential results as follows:
\bea
&&\!\!\!\!\!\!\!\!\Omega\left(T,\mu,\sigma,\Phi,\Phi^*\right)\!=\!\mathcal{U}
\left(\Phi,\Phi^*,T
\right)\!\!\!\!\!\!
\\
&&~~~~~~~~~~~~~~~~~~+\frac{\sigma^{2}}{2G}
-2N_f\int\frac{\mathrm{d}^3p}{\left(2\pi\right)^3}
\left\{3E_p\theta\left(
\Lambda^2-\vec{p}^2\right)
\right.
\nonumber\\
&&\!\!\!\!+\!\left.T\!\,\ln\!\left[1+3{\Phi}\mathrm{e}^{-\left(E_p-\mu\right)/T}\!\!+
3{\Phi^*}\mathrm{e}^{-{2}\left(E_p-\mu\right)/T}+
\!\!\mathrm{e}^{-3\left(E_p-\mu\right)/T}\right]\right.
\nonumber\\
&&\!\!\!\!+\!\left.T\! \,\ln\!\left[1\!+\!3{\Phi^*}\mathrm{e}^{-\left(E_p+\mu\right)/T}\!\!+\!
3{\Phi}\mathrm{e}^{-{2}\left(E_p+\mu\right)/T}\!\!+
\!\mathrm{e}^{-{3}\left(E_p+\mu\right)/T}\right]\right\}
\nonumber
\eea
where the quark quasiparticle energy is $E_p=\sqrt{\vec{p}^2+m^2}$ and the dynamical (constituent)
quark mass is the same as in the standard NJL model:
$m=m_0-\langle\sigma\rangle=m_0-G\langle\bar{\psi}\psi\rangle$.

In general, the fields $\Phi$  and $\Phi^*$ (and their thermal expectation values) are different at non-zero quark chemical potental~\cite{dpz05}. We demonstrate in a forthcoming paper~\cite{Simon}  that $\Phi\neq\Phi^*$ at $\mu \neq 0$ is primarily a consequence of quantum fluctuations of 
the fields around their mean field values. In the present work we stay at the mean-field limit which consistently implies $\Phi =\Phi^*$.

Minimizing $\Omega(T,~\mu,~\sigma,~\Phi)$ determines the chiral condensate $\sigma$ and the Polyakov loop $\Phi$ as
functions of temperature and quark chemical potential.
Fig.~\ref{fig2} shows the temperature dependence of the Polyakov loop at $\mu=0$ for two situations: first for the pure-gauge case (dotted line), and secondly for the case including quarks (full line). 
While the pure-gauge lattice results have been used to fix the effective potential for the Polyakov loop filed as explained previously, the case with inclusion of quarks is a prediction of the model without any additional tuning of parameters. The resulting agreement with corresponding lattice data incorporating dynamical quarks~\cite{Kaczmarek:2005} is striking. 

Notice that, while the Polyakov loop shows a 
discontinuity at the critical temperature in the pure gauge system 
indicating a first order deconfinement phase transition, this turns into a 
smooth crossover when quarks are introduced. At the same time
the critical temperature is reduced from $T_0 = 270$ MeV in the pure gauge case
to around $T_c=215$ MeV in the presence of quarks (not evident from 
Fig.~\ref{fig2} since the horizontal axis is scaled by $T_c$).
\begin{table}
\begin{center}
\begin{tabular}{|c|c|c|}
\hline
\hline
&&\vspace{-.3 cm}\\
$\Lambda$ [GeV]&$G$[GeV$^{-2}$]&$m_0$[MeV]\\
&&\vspace{-.3cm}\\
\hline
&&\vspace{-.3cm}\\
0.651&10.08&5.5\\
&&\vspace{-.3cm}\\
\hline
\hline
&&\vspace{-.3cm}\\
$|\langle{\bar \psi}_u\psi_u\rangle|$$^{1/3}$[MeV]&$f_{\pi}$[MeV]&$m_{\pi}$[MeV]\\
&&\vspace{-.3cm}\\
\hline
&&\vspace{-.3cm}\\
251&92.3&139.3\\
\hline
\hline
\end{tabular}
\caption{Parameter set used in this work for the NJL model part of the effective Lagrangian~(\ref{lagr}),
and the resulting physical quantities. For these values of the parameters we
obtain a constituent quark mass $m=$325 MeV.}
\label{tab2}
\end{center}
\end{table}
\begin{figure*}[t]
\begin{minipage}{.48\textwidth}
\resizebox{0.95\textwidth} {!}
{\includegraphics{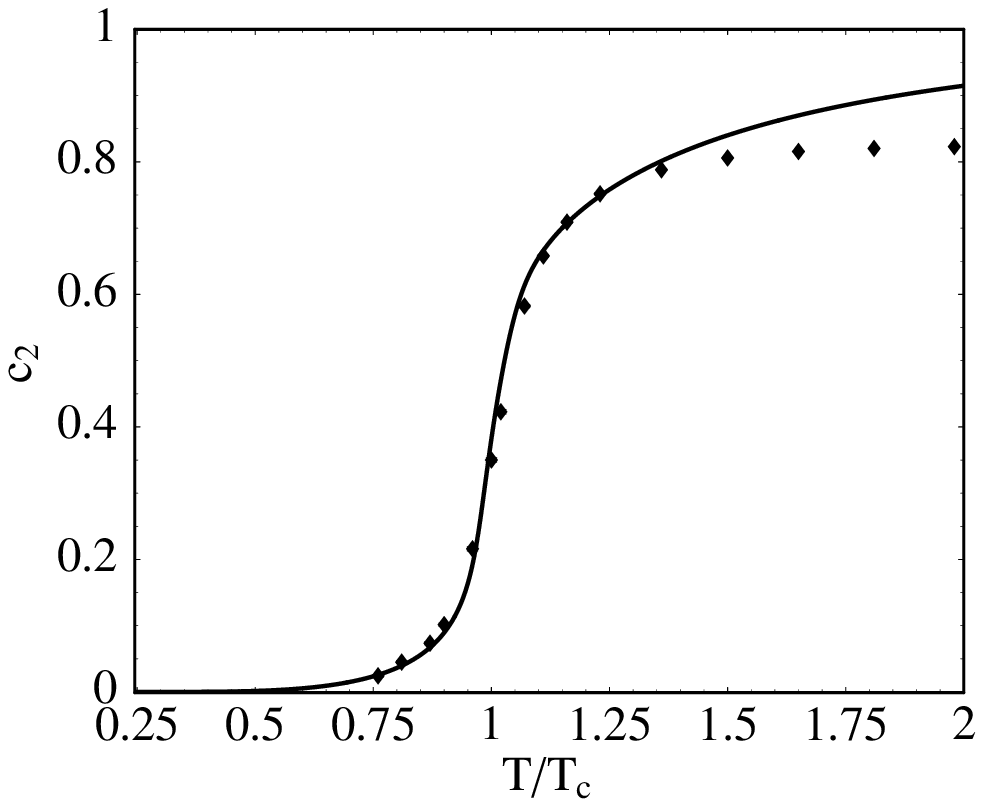}}
\end{minipage}
\begin{minipage}{.48\textwidth}
\resizebox{0.95\textwidth} {!}
{\includegraphics{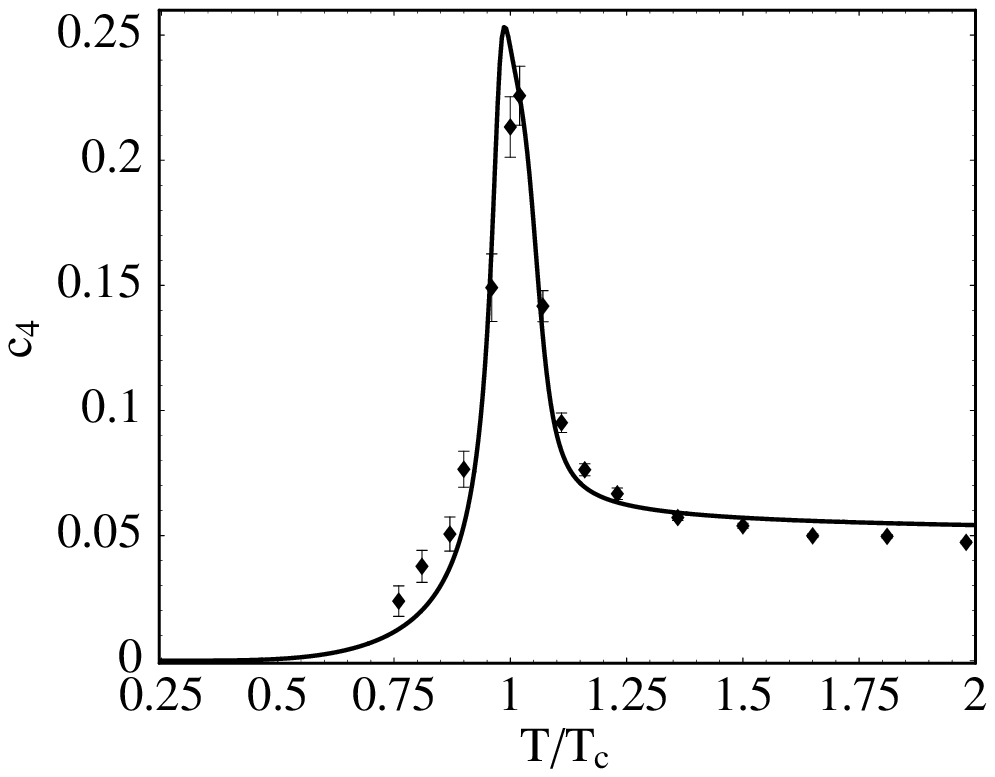}}
\end{minipage}
\caption{Expansion coefficients $c_2$ and $c_4$: the PNJL model
results are compared with the corresponding lattice results taken
from Ref.~\cite{Allton:2005gk}.} 
\label{fig3}       
\end{figure*}
\subsection{Comparison with lattice results}
In our previous work~\cite{Ratti1} we have compared predictions
of the PNJL model concerning QCD thermodynamics, at zero and finite
$\mu$, with corresponding lattice data. At non-zero
chemical potential the lattice ``data"
are deduced from a Taylor expansion of the thermodynamic quantities in powers of  
$\mu/T$ around $\mu = 0$. 
For this reason, following also Ref.~\cite{Ghosh:2006qh}, we perform 
the same expansion in the PNJL model, and compare the lattice expansion
coefficients to those calculated in our model.
We have:
\bea
\frac{p\left(T,\mu\right)}{T^4}&=&\sum_{n=0}^{\infty}c_n\left(T\right)
\left(\frac{\mu}{T}\right)^n;
\nonumber\\
c_n\left(T\right)
&=&\left.\frac{1}{n!}\frac{\partial\left(p\left(T,\mu\right)/T^4\right)}
{\partial\left(\mu/T
\right)^n}\right |_{\mu=0}
\label{pressure}
\eea
Our results for the coefficients $c_2$ and $c_4$ are shown in 
Fig.~\ref{fig3}. Once a limited set of input parameters
is fitted to Lattice QCD in the pure gauge sector and to pion properties in the
hadronic sector,
the PNJL model evidently provides a very good description of the lattice
data for the expansion coefficients of the Taylor series.
The results for $c_6$ (not shown here) are also in very good agreement with the lattice data and 
will be found in a forthcoming paper~\cite{Simon2}.

From these coefficients the Taylor-expanded 
pressure~(\ref{pressure}) is reconstructed and the quark number density follows as:
\bea
\frac{n_q}{T^3}=2c_2\left(\frac{\mu}{T}\right)+4c_4\left(\frac{\mu}{T}
\right)^3+6c_6\left(\frac{\mu}{T}\right)^5+...
\eea

%
\begin{figure*}
\begin{minipage}{.48\textwidth}
\resizebox{0.95\textwidth} {!}
{\includegraphics{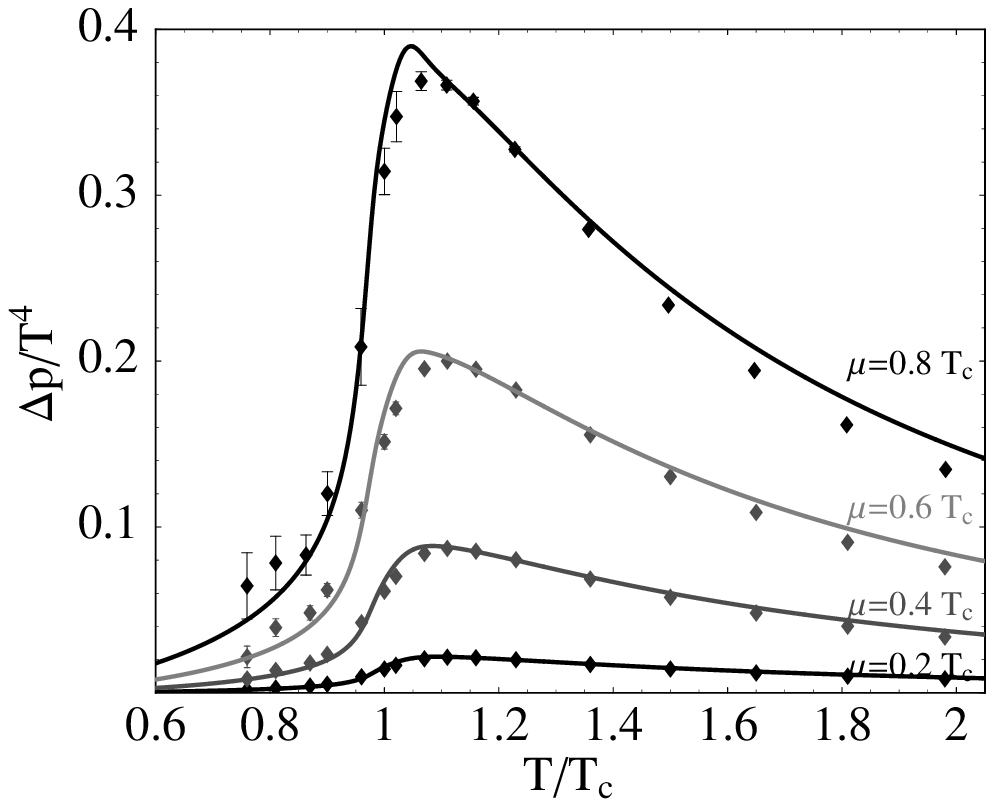}}
\end{minipage}
\begin{minipage}{.48\textwidth}
\resizebox{0.95\textwidth} {!}
{\includegraphics{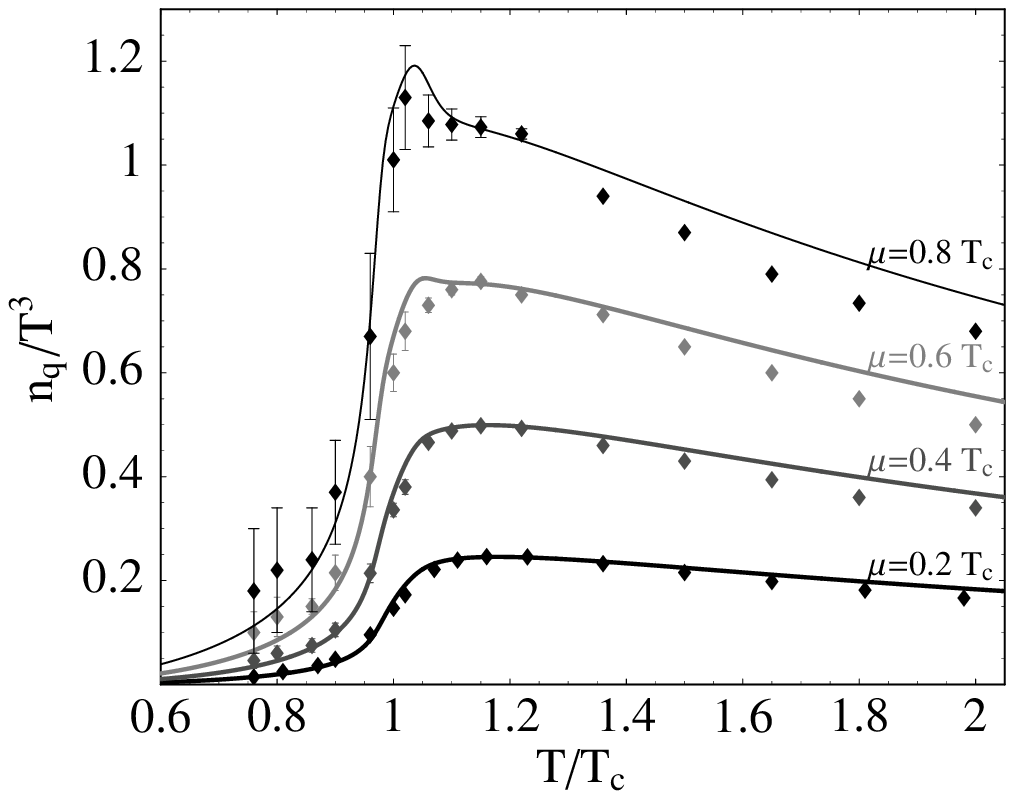}}
\end{minipage}
\caption{Scaled pressure difference (left) and quark number density (right) as 
functions of $T/T_c$ for different values of $\mu/T_c$. In both cases
the lattice data, obtained from a 
Taylor expansion to sixth order in quark chemical potential
~\cite{Allton:2005gk}, are
compared to the PNJL model results, also obtained from a Taylor expansion
up to sixth order in quark chemical potential.}

\label{fig5}       
\end{figure*}
\begin{figure*}
\begin{minipage}{.48\textwidth}
\resizebox{0.95\textwidth} {!}
{\includegraphics{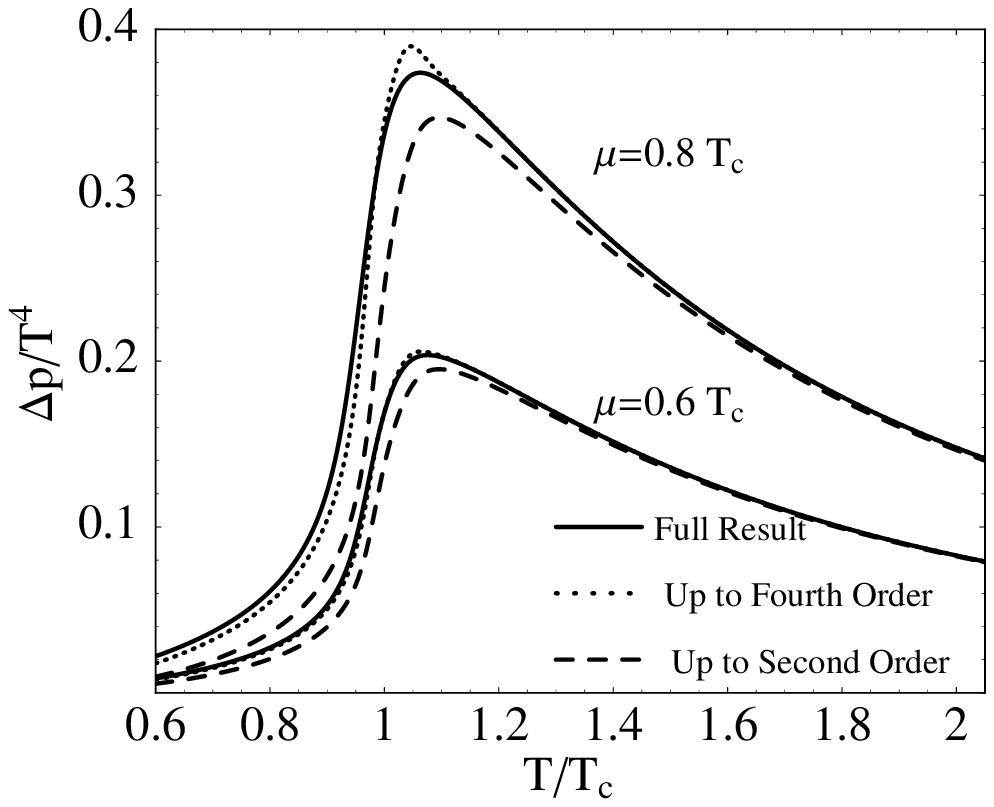}}
\end{minipage}
\begin{minipage}{.48\textwidth}
\resizebox{0.95\textwidth} {!}
{\includegraphics{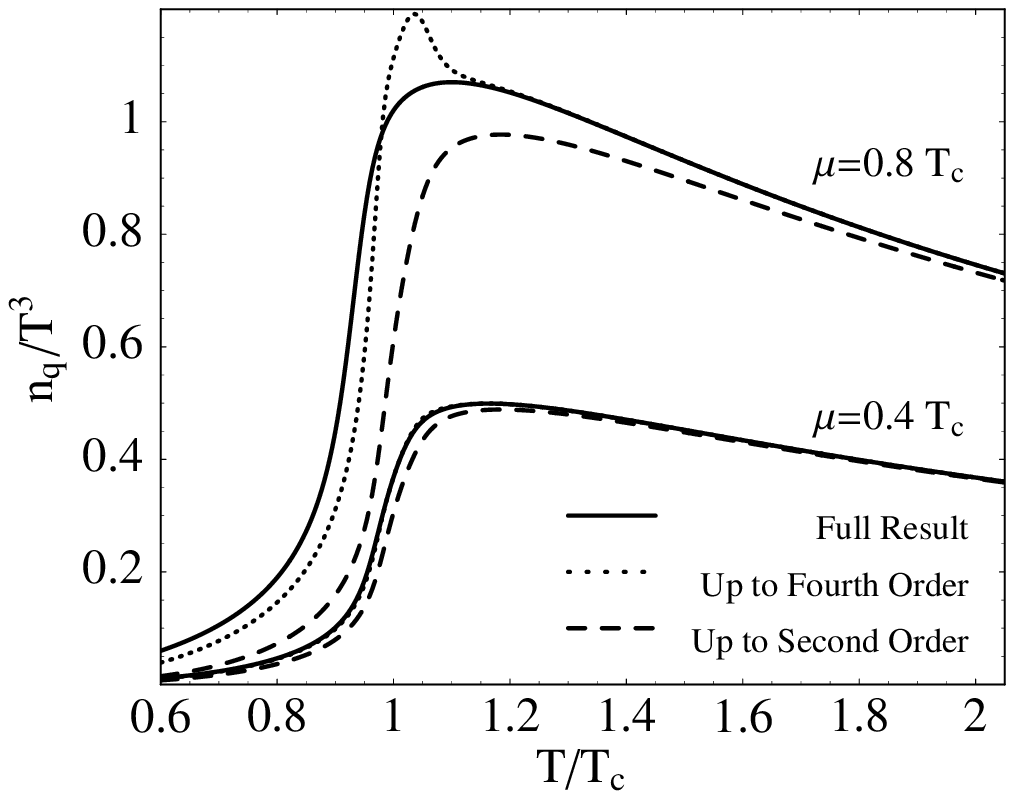}}
\end{minipage}
\caption{Pressure difference (left) and quark number density (right) as 
functions of $T/T_c$ for two different values of quark chemical potential.
The full result (continuous line) is compared to the Taylor expansion up
to second (dashed line) and fourth order (dotted line) in quark chemical
potential.}
\label{fig6}       
\end{figure*}

%
Fig.~\ref{fig5} displays results for the pressure difference, $\Delta p = 
p(T,\mu) - p(T,\mu=0)$, and for the
quark number density, both expanded around  $\mu=0$ up to sixth order in quark chemical potential, consistently with the corresponding lattice data. 
The full computation of the
pressure and quark number density at finite chemical potential cannot be 
performed on the lattice, but this is of course possible in our model.
We can thus investigate convergence properties of
the Taylor series and study discrepancies between the full and 
truncated results. This study is shown in Fig.~\ref{fig6}. The continuous lines (full results) are compared to second order (dashed) and fourth order (dotted) expansions in $\mu\over T$.  Evidently the series converges rapidly for small chemical potentials: the second order differs by 
less than $5\%$ from the full result, and the fourth order 
basically coincides with the solid curve. 
It is nevertheless also evident that when $\mu/T_c$ becomes
larger, of order 1, the Taylor expansion develops problems around $T_c$.
In this region in fact, the expansion coefficient $c_4$ is rather large, 
and for this reason the fourth order contribution overshoots the full one 
around $T_c$. When $T$ becomes larger, the coefficient $c_4$ drops
and for this reason the agreement between the full result and the 
truncated one is again very good.
\section{Conclusions}
The PNJL model represents a minimal synthesis of the two basic principles that 
govern QCD at low temperatures: spontaneous chiral symmetry breaking and 
confinement. The  respective order parameters (the chiral quark condensate and 
the Polyakov loop) are given the meaning of collective degrees of freedom. 
Quarks couple to these collective fields according to the symmetry rules 
dictated by QCD itself.
Once a limited set of input parameters is fitted to Lattice QCD in the pure 
gauge sector and to pion properties in the hadron sector, the quark-gluon 
thermodynamics above $T_c$ up to about twice the critical temperature is well 
reproduced. In particular, the coefficients of the Taylor expansion in $\mu/T$  are in remarkably good agreement with the ones coming from lattice simulations.
The PNJL model correctly describes the step from the 
first-order deconfinement transition observed in pure-gauge Lattice QCD to the 
crossover transition when $N_f = 2$ light quark flavours are added.
The predicted behaviour of the Polyakov loop as a function of temperature in the
presence of two quark flavours is in surprisingly
good agreement with the corresponding lattice data: a highly non-trivial result.

The Taylor-expanded pressure difference and quark number 
density of the system has been calculated up to sixth order in quark chemical potential.
The agreement with corresponding lattice data is again very good.
Finally, a comparison between the full result 
for these two quantities and the truncated one has been performed.
We observe fast convergence of the 
expansion in powers of $\mu/T$ at small chemical potential. At larger 
chemical potentials, comparable to $T_c$, limitations in the applicability of the Taylor expansion become apparent.

%
%

\end{document}